\newfont {\xx} {cmti10}

\newcommand{\bea}{\begin{eqnarray}}
\newcommand{\eea}{\end{eqnarray}}
\documentstyle[epsf,epsfig,12pt]{article}
\begin{document}
\pagestyle{empty}
\begin{flushright}
{CERN-TH/96-108}\\
{SHEP-96-10}\\
\end{flushright}
\vspace*{5mm}
\begin{center}
{\large {\bf Virtual-Sparticle Threshold Effects
on Large-$E_T$ Jet Cross Sections}}\\
\vspace*{1cm}
{\bf John Ellis}\\
\vspace{0.3cm}
Theoretical Physics Division, CERN \\
1211 Geneva 23, Switzerland\\
and \\
{\bf Douglas A. Ross} \\
\vspace{0.3cm}
Physics Department,\\
University of Southampton,\\
Southampton SO17 1BJ, United Kingdom\\
\vspace*{2cm}
{\bf ABSTRACT} \\ \end{center}
\vspace*{5mm}
\noindent
We discuss the one-loop virtual-sparticle corrections
to QCD jet cross sections at large $E_T$ and large dijet
invariant masses, with reference to present Tevatron and
future LHC collider experiments. We find characteristic
peaks and dips in the sparticle threshold region,
due to interferences with tree-level QCD diagrams. Their
magnitudes may be several per cent of the parton subprocess
cross sections, so they might provide a useful search tool
that is complementary to the usual missing-energy
signature for supersymmetry.

\vspace*{5cm}

\begin{flushleft}
CERN-TH/96-108\\
SHEP-96-10 \\
April 1996
\end{flushleft}
\vfill\eject

\setcounter{page}{1}
\pagestyle{plain}


Recent data from the CDF collaboration on large-$E_T$ hadronic
jets \cite{CDF} have stimulated interest in the possibility that jet
measurements at hadronic colliders may be sensitive to quantum
corrections due to virtual particles, such as those appearing
in supersymmetric models \cite{Betal,KW}.
Hadron colliders such as the CERN
$\bar p p$ collider, the Tevatron or the LHC are usually thought
of as exploratory machines, with precision physics left to
$e^+ e^-$ colliders such as LEP or the LHC. However, $e^+ e^-$
colliders are certainly also discovery machines, and the enormous
event rates at present and future hadron colliders may provide
precision tests of the Standard Model and its possible
extensions. The interpretation of large-$E_T$ jet cross sections
inherits uncertainties from the non-perturbative parton
distribution and fragmentation functions\footnote{It has
been suggested \cite{huston} that the
apparent discrepancy reported by CDF \cite{CDF}
may  be accommodated
by these uncertainties, particularly in the gluon distribution:
see however
\cite{GMRS}.
However, the CDF and D0 \cite{D0}
data need to be reconciled before this
issue can be resolved, after which one would know whether to
entertain speculations about new physics \cite{new},
of which compositeness \cite{CDF}
is not the most conservative.},
but effects on the shape of the cross section in the neighbourhood
of a new particle threshold may be discernible.

It is conceivable that precision jet
measurements could provide a useful new way to detect sparticles
indirectly. Even if they are being produced copiously,
their decays may be difficult to disentangle, e.g., if they
are dominated by cascades, or if R parity is violated. Thus,
the mass reach in direct searches could well be less than
the accessible range of $E_T$, as is now the case at the
Tevatron, and could in the future also be the case at the LHC.

The one-loop quantum corrections to jet cross sections in the
Standard Model have been calculated \cite{KEetal},
and studied extensively \cite{KS}.
First estimates of the one-loop virtual corrections in the
minimal supersymmetric extension of the Standard Model have
also been presented recently, including the ultraviolet
logarithms associated with the slower running of the strong
coupling $\alpha_s$ above the sparticle threshold \cite{Betal},
and sub-threshold effects \cite{KW}.
However, a complete calculation
which matches these contributions is not available, and the
infrared logarithms associated with the separation between
final states that do or do not contain sparticles have not
been considered. In view of the physics interest mentioned
in the previous paragraph, it is desirable
to calculate exactly the one-loop
sparticle effects from the threshold region on upwards.

In this paper, we present the main results of such a
calculation, including self-energy, vertex and relevant
aspects of ``box'' diagram contributions to the parton
subprocesses $\bar q q \rightarrow \bar q q,
q q \rightarrow q q, \bar q q \rightarrow g g$
and $q g \rightarrow q g$, which are expected to dominate large-$E_T$
cross sections at the Tevatron and LHC. We present
results which combine and match
the behaviours below and above threshold, displaying them
 graphically as
functions of the subprocess energy and centre-of-mass scattering
angle. The
numerical results we find are quite small, so that
measurements with a statistical accuracy of the order of a percent
in bins with widths of the order of $10 \%$ would be required to
see effects in the $E_T$ or dijet mass distribution.
A more
detailed presentation of our calculations, together with more
discussion of their observability based on
convolution with sample parton distribution functions, will
be presented in a subsequent paper \cite{ERmore}.


We work in the context of the Minimal Supersymmetric extension of
the Standard Model (MSSM), assuming for simplicity that the
squark partners of both the left- and right-handed helicity
states ${\tilde q}_{L,R}$
of the five lightest quark flavours are degenerate with
common mass $m$. The gluino mass we denote by $M$. For
simplicity, and to maximize the possible threshold effects,
we assume equal masses for the squarks and gluinos, i.e.  $m = M$.
If the masses are very different,
most of the higher-order contributions
are suppressed below the higher-mass threshold. The effects of moderate
mass differences between squarks and gluinos
will be discussed in~\cite{ERmore}. We consider radiative corrections to
the scattering and production of gluons and the five lightest
quark flavours, recognizing that $\bar t t$ production
requires a separate treatment~\cite{susyt}.

\begin{figure}
\begin{center}
\leavevmode
\hbox{\epsfxsize=4.0 in
\epsfbox{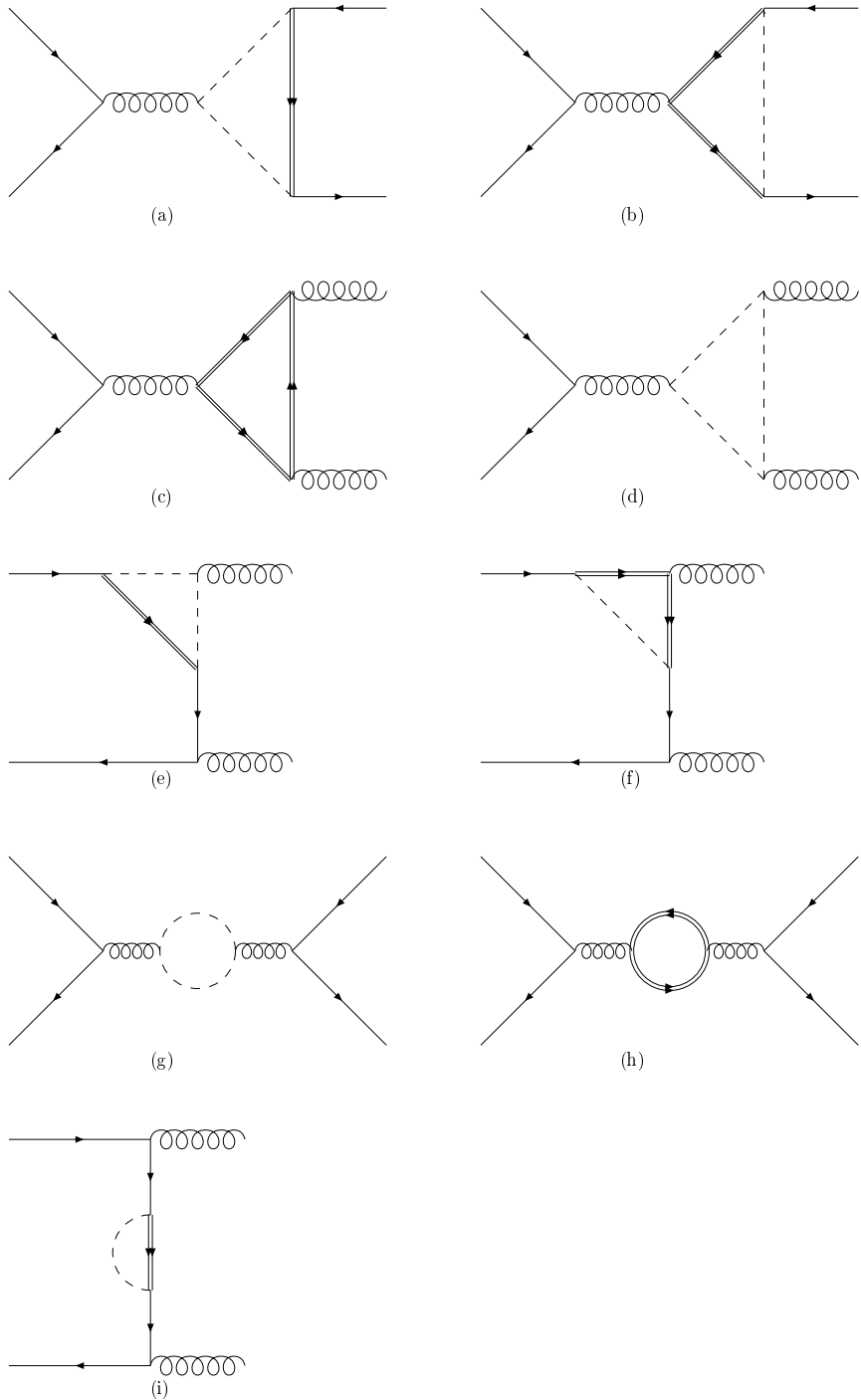}}
\end{center} \vspace*{-.5in}
\caption{One-loop Feynman diagrams involving virtual
sparticles in the MSSM for
(a), (b) the $g \bar q q$ vertex in $q_i {\bar q}_i
\rightarrow q_j {\bar q}_j$ scattering,
(c), (d) the $g g g$ vertex in $q {\bar q} \rightarrow g g$
scattering,
(e), (f) the $g \bar q q$ vertex in $q {\bar q} \rightarrow
g g$ scattering,
(g), (h) the gluon self energy,
and (i) the quark self energy. The broken lines represent squarks
and the double solid lines represent gluinos.} \label{fig1}
\end{figure}

In this paper, we restrict our calculations to the contributions
of self-energy insertions and vertex corrections shown in Fig.~1,
which display the most dramatic behaviours in the theshold
region. Previous experience with higher-order corrections
indicates that contributions from ``box'' diagrams are likely to be
small in the threshold region. Moreover,
the results of ref. \cite{KW} indicate
that the ``box''-diagram contributions are significantly smaller than
vertex and self-energy contributions,
also at energies well below threshold.
Nevertheless, the effects of these ``boxes'' will be included for
completeness in ref. \cite{ERmore}.

In order to match our higher-order calculation to a leading-order
calculation in terms of a value of $\alpha_s$ extracted
from data assuming the absence of sparticle loops, we use
a renormalisation prescription in which the contributions
for loops containing sparticles vanish in the low-momentum
limit. In the case where the masses of the sparticles are all taken to be
equal, this turns out to be equivalent to the
$\overline{MS}$ scheme with
renormalisation mass scale $\mu$ set
equal to the sparticle mass $M \ (=m)$.

The fact that we perform a complete calculation extending
from far below the sparticle threshold to very high energies
provides us with several checks on the relative magnitudes
and signs of the various diagrams shown in Fig.~1.
In particular, the fact that the contribution to the $\beta$ function
from sparticle loops arises entirely from their contributions to the
gluon self-energy provides us with the check that the non-abelian parts
of the ultraviolet divergences of Figs 1(a) and 1(b) cancel
against each other, and likewise
Figs. 1(e) and 1(f). Moreover, for the same reason the ultraviolet divergence
of Fig. 1(c) cancels against that of Fig 1(h), and likewise
Fig 1(d) against  1(g). Furthermore, the abelian Ward-Takahashi identity tells
 us that
the abelian part of the sum of the ultraviolet divergences in Figs 1(a) and
 1(b)
(and likewise Figs. 1(e) and 1(f)) cancel against the ultraviolet-divergent part
 of
the fermion self energy of Fig. 1(i).
 We have also extracted the low-energy
expansions $\propto q^2/m^2$ of our results, for comparison
with \cite{KW}. Despite some minor differences, we confirm
the general magnitude of the below-threshold corrections
found in \cite{KW}.

\begin{figure}
\hbox to \hsize{\hss
\epsfxsize=0.4\hsize
\epsffile{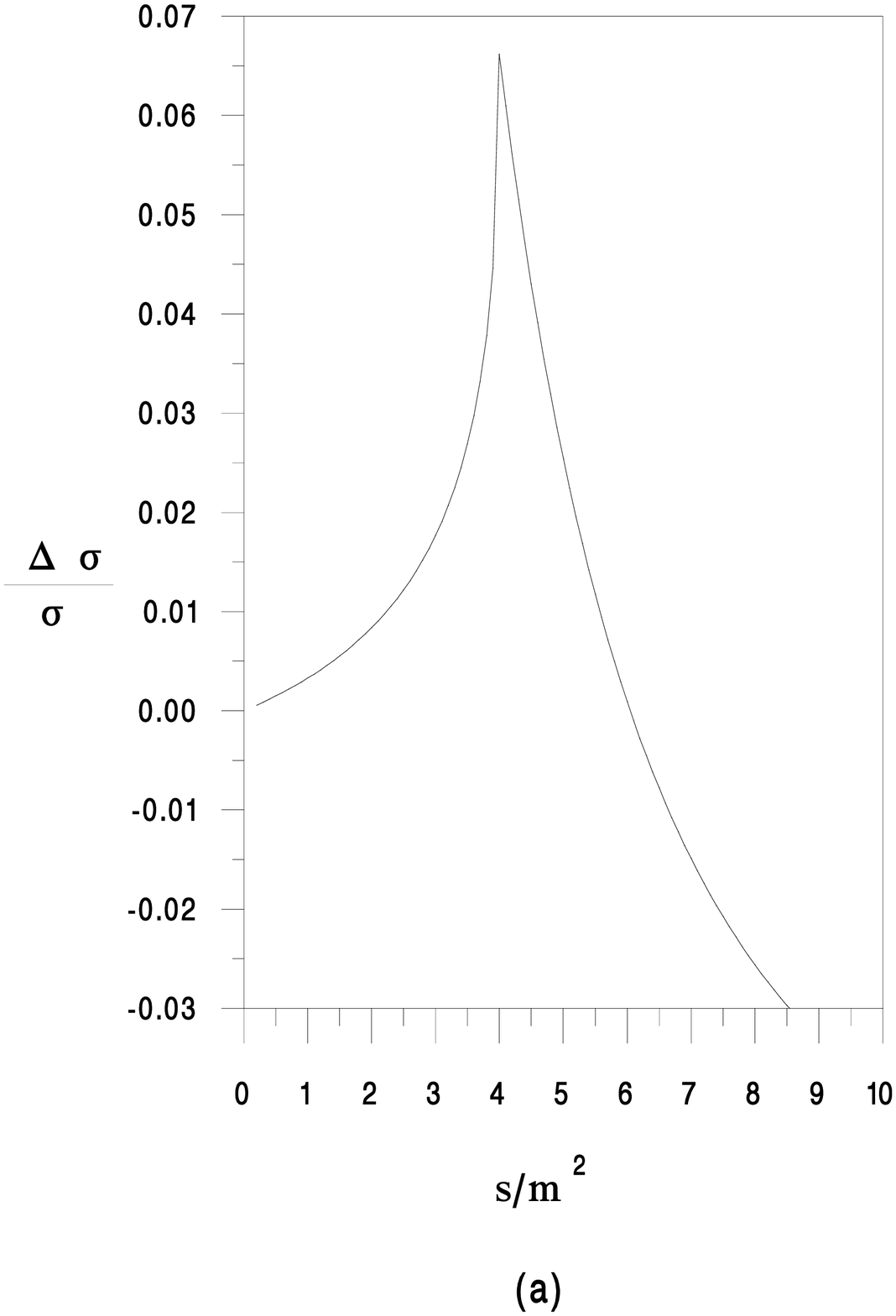}
\hfill
\epsfxsize=0.4\hsize
\epsffile{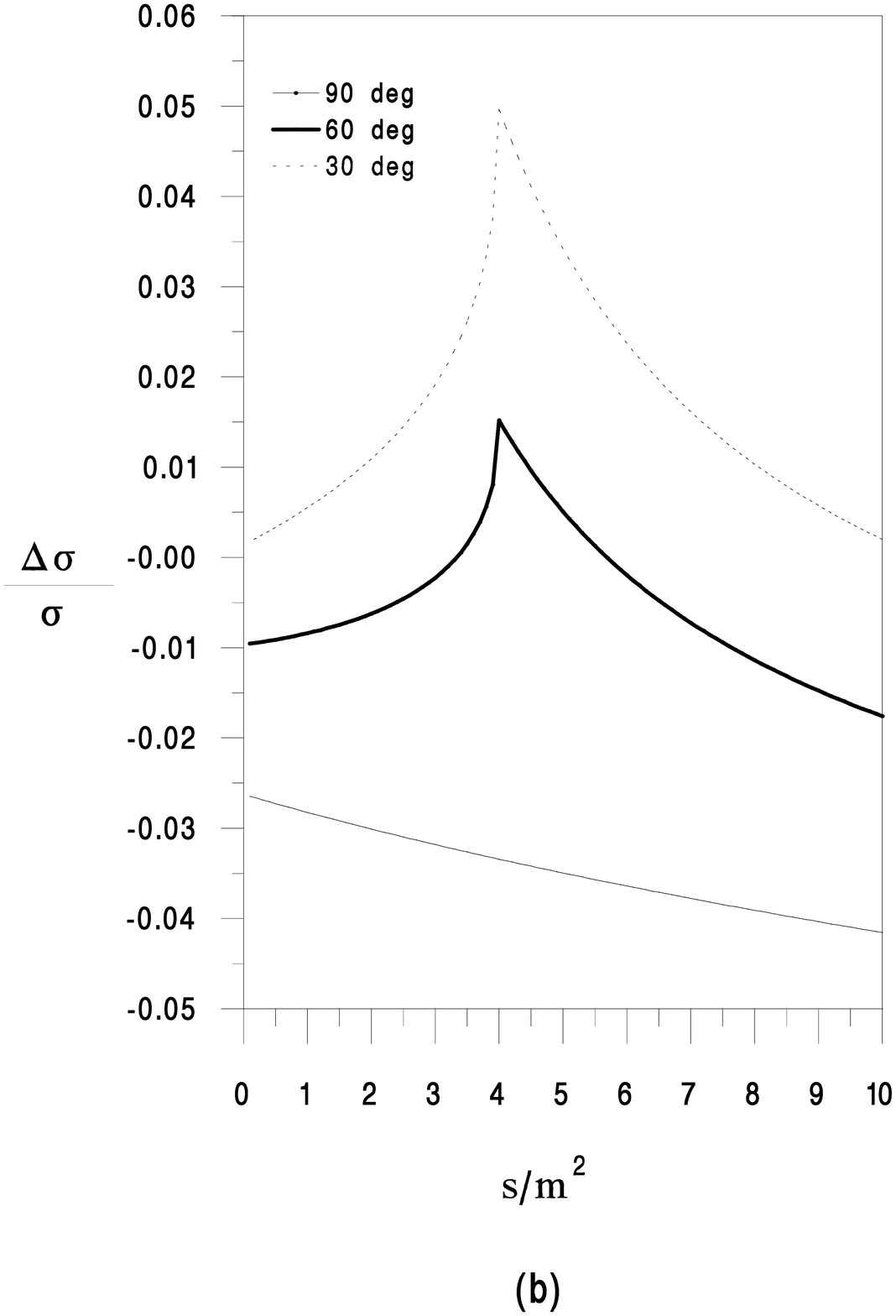}
\hss}
\kern2em
\hbox to \hsize{\hss
\epsfxsize=0.4\hsize
\epsffile{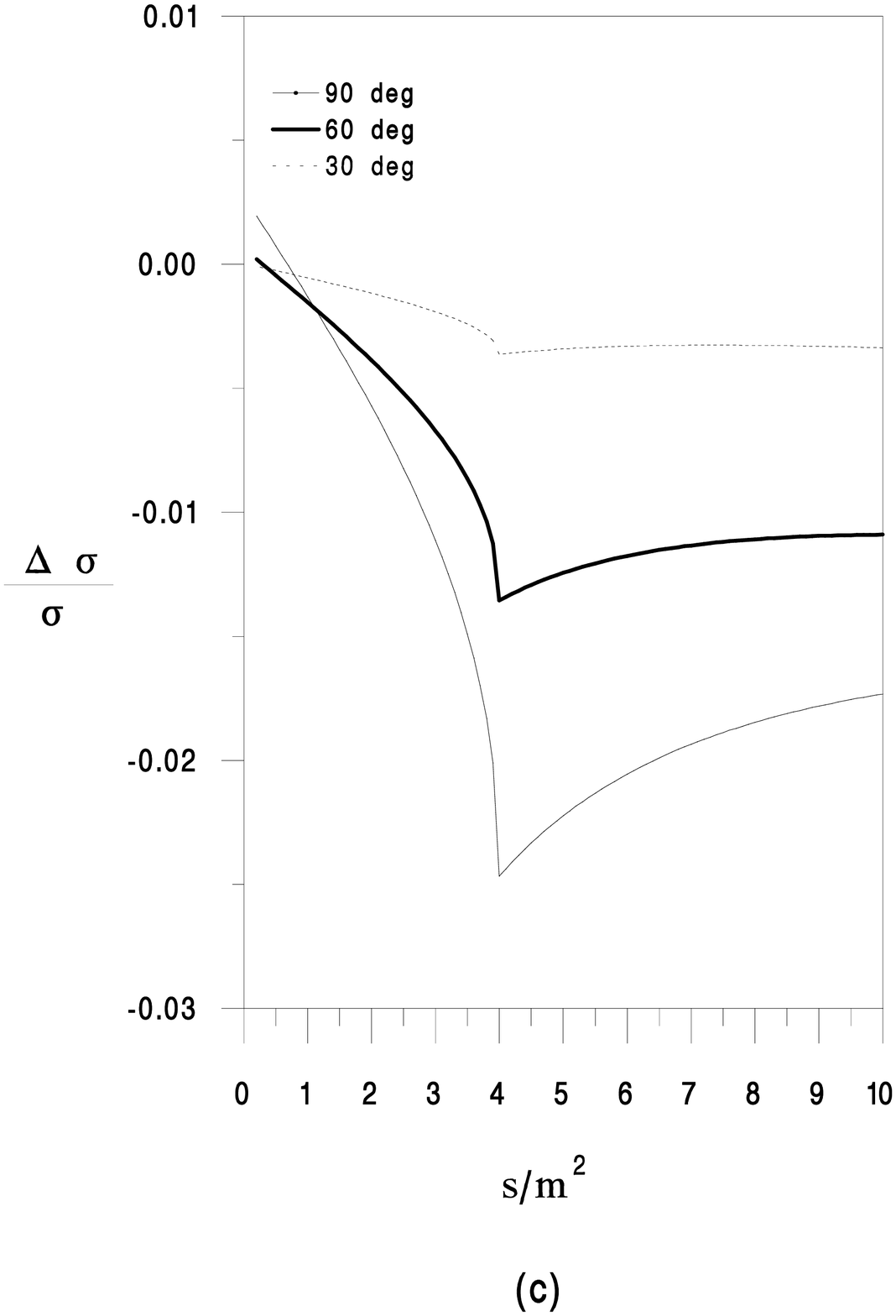}
\hfill
\epsfxsize=0.4\hsize
\epsffile{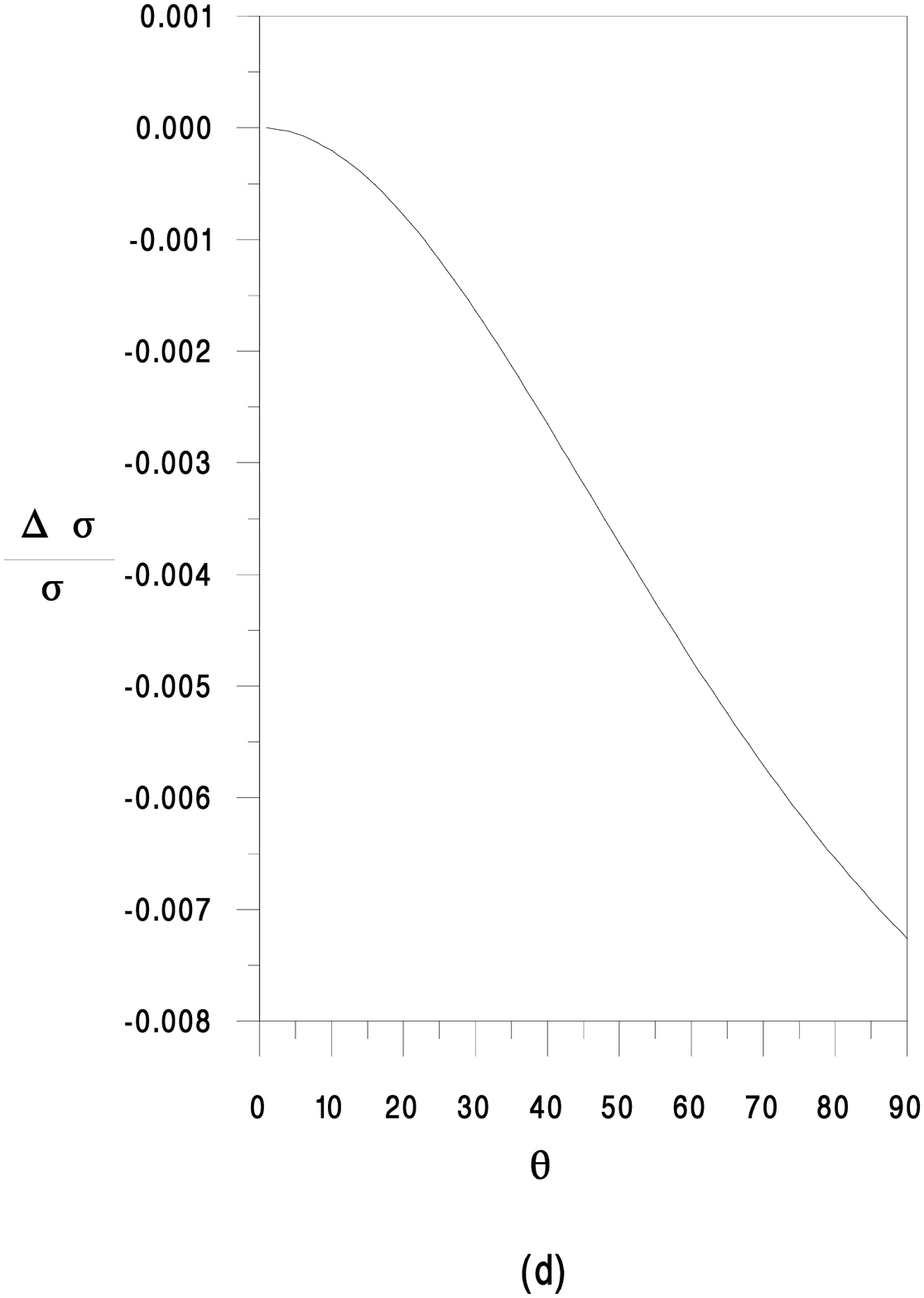}
\hss}
\caption{One-loop virtual-sparticle corrections in the threshold
region of the subprocess centre-of-mass energy squared $s$
to the processes (a)
$q_j \, \bar{q}_j \rightarrow q_k \, \bar{q}_k$,
 (b) $q  \, \bar{q} \rightarrow g \, g$ for
three different subprocess centre-of-mass scattering angles,
(c) $q \, g \rightarrow  q \, g$
also for three different values of scattering angle, and (d)
 $q_j \, q_k \rightarrow
q_j \, q_k$, against the centre-of-mass subprocess
scattering angle, $\theta$, for $s=10m^2$.
All corrections are evaluated
using $\alpha_s=0.11$.}
\label{fig2}
\end{figure}

Our main results are shown in Fig.~\ref{fig2},
 in which we plot the
one-loop virtual-sparticle correction to the subprocesses
(a)  $q_j \, \bar{q}_j \rightarrow q_k \, \bar{q}_k$, (b)
$q  \, \bar{q} \rightarrow g \, g$,    and (c)
$q \, g         \rightarrow  q \, g $,
as  functions of the square of the subprocess
centre-of-mass energy, $s$.
Full details of the results shown in these
plots, as well as others,
will be presented in \cite{ERmore}. In this
letter, we focus your attention on the most
important features of these plots,
namely the distinctive peak and dip structures
in the correction
at the threshold $s=4m^2$. These are consequences of the familiar
discontinuity of the derivative of the real part of any amplitude
at the branch point where the amplitude develops an imaginary part.
Since the order $\alpha_s$ corrections are due to the interference
between the tree and one-loop amplitudes, one cannot be sure that
the effect will be positive. Indeed, the correction to the
$q \, g \rightarrow q \, g$ subprocess
shown in Fig.~\ref{fig2} (c)  is negative.

The correction to the process  $q \, \bar{q} \rightarrow q \, \bar{q} $
shown in Fig.~2(a) depends only  on $s$, whereas
the corrections to the other processes are also functions of
the squared subprocess momentum transfer $t$.
We have plotted in Figs.~2(b) and (c) the threshold behaviours
at three different values of the centre-of-mass
scattering angle.
The fact that the threshold behaviour has a characteristic
angular dependence
means that it should be possible in principle to
distinguish such virtual-sparticle threshold effects
from other possible mechanisms for a structure in the dijet
invariant mass, such as the production of a new particle
with definite spin. Also shown
in Fig.~\ref{fig2}(d)  are the one-loop corrections to the process
$q_j \, q_k \rightarrow q_j \, q_k$, which is controlled by a gluon
exchange in the $t$ (or $u$) channel. Because of this, and
because its direct-channel quantum numbers are exotic from
the point of view of the MSSM, its corrections have
no distinctive threshold behaviour in $s$.
Therefore we have plotted it as a function of
scattering angle for one large
value of $s \ (=10m^2)$: we see that the corrections
 are negligible compared
with those to the other subprocesses.

 The one-loop virtual-sparticle corrections to the subprocess
$q_i {\bar q}_j \rightarrow q_i {\bar q}_j$ are identical to
those for $q_i q_j \rightarrow q_i q_j$. Since this cross-channel
exchange reaction dominates over $q_i {\bar q}_i$ annihilation,
the threshold effect shown in Fig.~2~(a) will be strongly
diluted if one studies the dijet mass distribution without
any flavour separation. We have also calculated virtual-sparticle
threshold effects in the subprocess $g g \rightarrow g g$, but
do not display the results here, as elastic $g g$ scattering
is relatively
significant only at low values of the dijet mass, below those
of interest at the Tevatron. We find for this subprocess
a substantial threshold enhancement comparable in size and
width to that in Fig.~2~(a) \cite{ERmore}.

 Knowledge of the threshold structures in Figs.~2 (a),(b),(c)
supplements
the corrections in the low-energy limit $s \ll m^2$
calculated in
\cite{KW}. There it was estimated
that internal sparticle effects are likely to be of order 1\%, and
we find here that corrections as large as 5-6 \% can be obtained
in some subprocesses, albeit in a narrow energy region
around the theshold.

The suggestion \cite{Betal} that the effects of internal sparticle
loops can be estimated from their effect on the running of the coupling
must also be treated with caution. In the first place,
the running of the coupling only gives information about leading
logarithms, which would dominate the corrections only at energies far
above the threshold, and {\it cannot} be used to estimate the
correction in the threshold region\footnote{At energies sufficiently
far above threshold, one should also take into account sparticle
effects on the evolution of the parton structure functions, but
this effect is not relevant in the threshold region that we study
here.}. Furthermore, one can only extract
leading logarithms from the behaviour of the running of the coupling
for totally inclusive processes.  More precisely, the leading
logarithms can be obtained from the running of the coupling only
for infrared-safe processes in which there are no other relevant mass
scales. In the case under consideration, there are contributions
which would give rise to  collinear divergences
in the limit where the sparticle masses were taken to be zero.
Such divergences would normally be cancelled by the emission of real
 collinear sparticles if these were indistinguishable from
final states containing final state sparticles.
In the real world, however, the emission of massive
sparticles would have a clear  missing-energy (if $R$ parity is
conserved) or multi-jet/lepton (if $R$ parity is not conserved)
signal. Therefore it is possible
to distinguish clearly between the real and virtual sparticle
corrections, each of which has (equal and opposite) large logarithms
in the high-energy limit, over and above those extracted form the
running of the coupling. These logarithms only dominate the
corrections to the cross-sections at energies much larger
than those displayed in Fig. \ref{fig2}. In fact we have checked
numerically that they only set in for values of $s$ greater than
$1000 m^2$ in the case of quark production and
$s \, > \, 100 m^2$ for gluon production.

The structures shown in Figs. \ref{fig2} (a),(b),(c) are rather sharp. In order
to determine their effect on the $E_T$ cross-sections, it is necessary
to perform a convolution of the cross section for each subprocess with
the corresponding parton distribution functions \cite{KS}.
We defer a full discussion
of this to ref.\cite{ERmore}, but it is clear that
there will be some structure in the $E_T$  distribution around
$E_T=m$ analogous to that at $M_W/2$ in  the
charged-lepton energy spectrum in $W \, \rightarrow \ell \, \nu$
  decay.
Quantification of this effect requires a convolution of the
subprocess cross sections calculated here with the
appropriate parton distributions, which is beyond the scope
of this work. However,
jet $E_T$ distributions are probably
not ideal quantities for observing this
particular signal for supersymmetry.

It would be preferable to study experimentally
the distribution in the dijet mass
$M_X$, which would exhibit directly the threshold structure at
$M_X=2m$.
Since the fractional experimental resolution in this
quantity improves at higher energies, it may be possible
for measurements at the LHC to
achieve a resolution which is comparable with
the widths of the peaks shown
in Fig.~\ref{fig2}. It appears possible that the LHC experiments
might be able to achieve a resolution in the dijet invariant
mass of around $10$ \%. This is comparable to the widths of the
peaks shown in Fig.~2.  We do not address here the
convolution of the parton subprocesses we calculate with
realistic parton distributions relevant to the Tevatron and LHC
experiments. However, their dijet mass resolutions
might be sufficient to offer
the possibility of observing an abrupt increase (or decrease) in a
subprocess
differential cross-section with respect to $M_X$ at $M_X=m$,
which could be
a very clean signal for the sparticle threshold.  If it could be
measured, the magnitude of the effect -
which is expected to of the order of 5 \% according
to the calculation reported here - would
be a valuable check that the new threshold was due to supersymmetry,
with the correct dynamics and magnitudes for the couplings of
sparticles to each other and to quarks and gluons.

\end{document}